# First Principles Study of Adsorption, Diffusion and Dissociation of NH$_3$ on Ni and Pd Surfaces.

Sergey Stolbov and Talat S. Rahman

Department of Physics, Cardwell Hall, Kansas State University
Manhattan, Kansas 55606, USA

**Abstract** Using the plane wave pseudopotential method within the density functional theory with the generalized gradient approximation for exchange and correlation potential, we have calculated adsorption energies ($E_{ad}$), diffusion barriers and the first dissociation barriers ($E_1$) for NH$_3$ on the Ni(111), Pd(111) and Ni(211) surfaces. The top sites are found to be preferred for NH$_3$ adsorption on Ni(111) and Pd(111). The diffusion barrier is calculated to be substantially higher for Pd(111) than for Ni(111). We also find that during the first dissociation step (NH$_3$ => NH$_2$ +H) on Ni(111) surface NH$_2$ moves from the top site to the nearest hollow site, while on Ni(211) it moves from the initial top site at the step edge to the bridge site in the same step chain. H is found to occupy the hollow sites for both surfaces. For the reaction on Ni(111), the $E_{ad}$ is found to be 0.23 eV lower than $E_1$, while at the step of Ni(211), $E_1$ and $E_{ad}$ are almost equal to each other. This suggests that the molecule will rather desorb on Ni(111) than dissociate, whereas at the step the dissociation is favorable.

## Introduction

Recent world-wide commitment to the development of hydrogen economy has generated enormous interest in searching sources and carriers of clean hydrogen that can be used in fuel cells. Hydrogen is currently produced on an industrial scale through steam reforming of natural gas. Coal and biomass are also considered as resources for production of hydrogen fuel [1]. The main disadvantage of these sources of hydrogen is that they (especially coal and biomass) generate a large amount of CO$_x$ as byproducts. Proton exchange membrane fuel cells and, especially, alkaline fuel cells (highly efficient low temperature fuel cells) are not tolerable to CO$_x$ and require CO$_x$-free hydrogen, whereas reduction of CO$_x$ admixture in hydrogen to a sufficient level is a complex problem [1]. A promising way to solve this problem is to use ammonia as hydrogen source and, in the same time, as hydrogen storage. Ammonia contains 17.8 wt% hydrogen and stores 30% more energy by liquid volume than liquid hydrogen. Only 16% of the energy stored in NH$_3$ is needed for its conversion to N$_2$ and H$_2$. Ammonia is the second largest synthetic commodity product of the world chemical industry and the infrastructure for ammonia transportation, distribution, storage and utilization is well-established [2]. It is also very important that NH$_3$ molecules are carbon free: NH$_3$ decomposition byproduct is only N$_2$ that is quite benign to the fuel cells. In addition, it does not produce environmental pollution. There is thus no need for purification of hydrogen produced by NH$_3$ decomposition. Ammonia can be used as on-board carrier of hydrogen where the latter can be released by catalytic NH$_3$ decomposition. This stage, however, requires efficient and inexpensive catalysts for NH$_3$ dissociation.

The catalytic decomposition of ammonia on various metals has been studied for three decades [3-10]. Currently the best catalyst for this reaction is found to be ruthenium. Comparison made for various metals provides the following sequence of the decomposition rates: Ru > Ni > Rh > Co > Ir > Fe >> Pt > Cr > Pd > Cu >> Te [5]. It is widely accepted that the reaction has a sequential character with the following steps: $2NH_3(gas) \Rightarrow 2NH_3(ad) \Rightarrow 2NH_2(ad) + 2H(ad) \Rightarrow 2NH(ad) + 4H(ad) \Rightarrow 2N(ad) + 6H(ad) \Rightarrow N_2(ad) + 3H_2(ad) \Rightarrow N_2(gas) + 3H_2(gas)$, where (gas) stands for gas phase and (ad) denotes adsorbed state. The mechanisms of these steps are studied both experimentally and theoretically for few metal surfaces (mostly for Ru) [5-19] have led to the conclusion that the rate limiting step in catalytic ammonia decomposition is the recombinative desorption of nitrogen: $2N(ad) \Rightarrow N_2(gas)$ [7,10,15-19]. There is however another obstacle for the reaction, namely, the relative strength of the $NH_3$ adsorption energy, $E_{ad}$, and the energy barrier, $E_1$, for the first step of dehydrogenation: $NH_3 \Rightarrow NH_2+H$. For several transition metals $E_{ad}$ and $E_1$ are found to have close values [7,8,15] which makes the dissociation and desorption of $NH_3$ highly competitive. Diffusion of $NH_3$ on the surface is also important process, since it is shown [7,20] that, if $NH_3$ adsorbs on a terrace, it first diffuses to a step or a local defect and then dissociates. This is why in this paper our focus is on adsorption, diffusion and the first dissociation step of $NH_3$ on metal surface. The sequence of the reaction rates for eleven catalysts [5] has led us to the choice of metals for our study. Although Ni and Pd have the same number of valence *d*-electrons, their reactivity for the $NH_3$ decomposition is found to be dramatically different and an attempt to find a correlation between these rates and any model parameter characterizing the reactivity fails [5]. Rationalization of the observed huge difference in the reactivity for these two apparently similar metals may help understand the fundamental reaction mechanisms This work is our first step toward this goal. Using the plane wave pseudopotential (PWPP) method [21] we calculate the adsorption energies and the diffusion activiation energy for $NH_3$ on the Ni(111) and Pd(111), as well as the energetics of the $NH_3 \Rightarrow NH_2 +H$ dissociation on Ni(111) and Ni(211), which is regularly stepped surface. We find the $E_{ad}/E_1$ ratio to be an important parameter that may control efficiency of catalysts for the $NH_3$ decomposition. Below we provide some details of the calculations which is followed by a summary of our results and discussion.

**Computational Details**

Calculations presented in this paper are performed within the density functional theory with the generalized-gradient approximation (GGA) for the exchange-correlation functional [22] using PWPP method [21]. To meet translation symmetry requirement we apply the slab approximation. For the (111) surfaces we use a supercell containing 5 metal layer slab and 12 Å vacuum. To reduce interaction between $NH_3$ adsorbed on the surface, the (2x2) geometry is chosen for the supercell along the surface that corresponds to four metal surface atoms per $NH_3$ molecule. The supercell for Ni(211) surface comprises of a 11 layer slab and 12 Å vacuum and is doubled along the step corresponding to six metal surface atoms per one $NH_3$ molecule. The total number of Ni atoms in the (211) supercell are 22.

The ultrasoft pseudopotentials [23] are used for all atoms under consideration. To obtain accurate energetics, we set cutoff energies for the plane-wave expansion of 350 eV for all systems. A Monkhorst-Pack $k$-point mesh is applied to sample the Brillouin zone [24]. We use (5x5x1) and (5x4x1) $k$-point samplings for the (111) and (211) surfaces, respectively. For the calculation of adsorption energies we involve all atoms in the system in relaxation that is done using the conjugated gradient method. The structures are relaxed until the forces acting on each atom converged better than 0.01 eV/Å. The adsorption energy is calculated for $NH_3$ as fallows:

$$E_{ad} = E_{tot}(NH_3/MS) - E_{tot}(NH_3) - E_{tot}(MS),$$

where $E_{tot}(NH_3/MS)$ is the total energy of the metal slab with $NH_3$ adsorbed on it, $E_{tot}(NH_3)$ is the total energy of $NH_3$ and $E_{tot}(MS)$ is the total energy of a clean metal slab. For all three system the supercells have the same size and shape.

For calculations of diffusion and dissociation paths, some degree of freedom of atoms explicitly involved in the process are frozen and the ionic relaxation is stopped at the 1 meV total energy convergence.

**Results and Discussion**

We have calculated $E_{ad}$ for $NH_3$ adsorbed on the top, fcc and hcp hollow sites of Ni(11) and Pd(111), as well as on the top site of the step chain in Ni(211). The calculation results are listed in Tab. 1. We find the top site is preferred for $NH_3$ adsorption on both Ni(111) and Pd(111). Calculations performed for $NH_3$ adsorption on Au(111) [25] and on Rh(111) [26] also show that the top site is preferred. The fact that all these metals have quite different electronic structure, but the same preferred adsorption site suggests that the nature of $NH_3$ adsorption is controlled rather by the local surface geometry than the surface electronic structure.

**Table 1. Adsorption energies of $NH_3$ on Ni and Pd surfaces.**

| Site | Ni | | | | Pd(111) | |
|---|---|---|---|---|---|---|
| | (111)-top | (111)-fcc | (111)-hcp | (211)-step | top | fcc |
| $E_{ad}$(eV) | -0.83 | -0.39 | -0.39 | -0.94 | -0.68 | -0.47 |

We have also calculated the activation barriers for diffusion of $NH_3$ on Ni(111) and Pd(111) from one top site to another via the bridge site. To calculate the energy profile we displaced $NH_3$ step-wise along the path, fixed N coordinates along the surface and allowed the rest of the system to relax. The 10 step energy profiles result in symmetric curves giving the energy barriers of 0.32 eV for Ni(111) and 0.40 eV for Pd(111). As mentioned above, experimentally $NH_3$ decomposition rate on Pd is found to be much lower than that on Ni [5]. Since diffusion of $NH_3$ on the catalyst surface is expected to be an important step in the reaction [7], the difference in the diffusion activation barriers obtained from our calculations may partially explain the difference in the obtained decomposition rates.

To model $NH_3 \Rightarrow NH_2+H$ dissociation on Ni(111), we start from initial state (IS) which is $NH_3$ adsorbed on a top site of the surface. The final state (FS) for H is found to be the hollow site closest to the molecule. To find FS for $NH_2$ we substantially increase the bond length between N and the dissociating H by moving the H toward its FS (green arrow in the left panel of Fig. 1). We fix the H position and let the rest of the system relax. We find that $NH_2$ spontaneously moves in the opposite direction from the top site to take the nearest hollow site position, as shown in the right panel of Fig. 1.

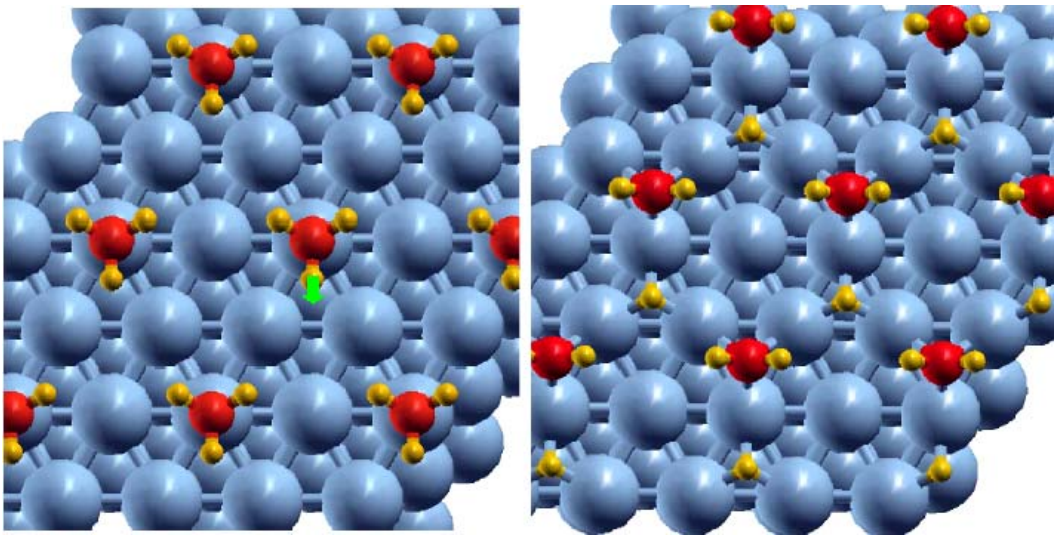

Fig. 1. Initial and final states for the $NH_3 \Rightarrow NH_2+H$ dissociation on Ni(111).

The final state for $NH_3$ dissociation on Ni(211) is found in the same manner with initially $NH_3$ adsorbed on the top of a step atom (see the left panel of Fig. 2). On this stepped Ni surface the FS for H is the hollow site closest to the step chain and that for $NH_2$ appears to be the bridge site in the step chain (see the right panel of Fig. 2).

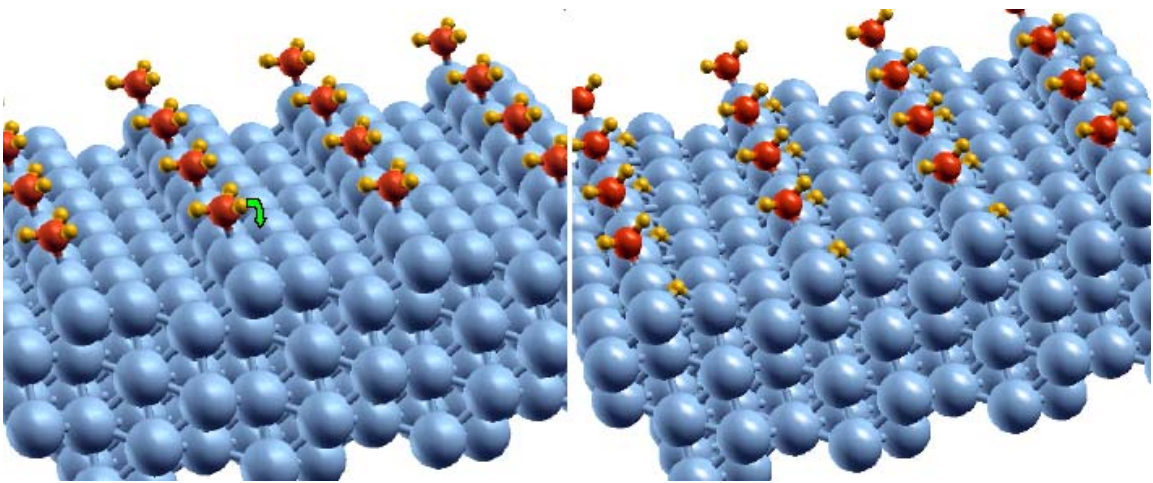

Fig. 2. Initial and final states for the $NH_3 \Rightarrow NH_2+H$ dissociation on the step of Ni(211).

Once the IS and FS for the reaction are defined, the dissociation barrier is calculated. In this preliminary study the optimal reaction path is found by grid method. In the future we propose to use more sophisticated methods that are available for the purpose. We find that $E_1$ is lower, if we increasing the bond between N and the dissociating H, tilt this bond simultaneously toward the surface.

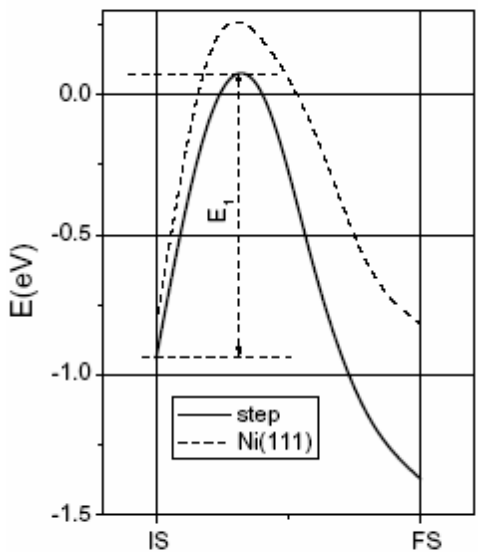

Fig. 7. The $NH_3 \Rightarrow NH_2 + H$ dissociation barriers calculated for the molecule adsorbed on the top site of the step at Ni(211) and on the top site of Ni(111) (terrace).

The energy profiles calculated along such dissociation paths for Ni(211) and Ni(111) are shown in Fig. 3. The FS energy is found to be much lower for dissociation at the step of Ni(211) than on Ni(111). The $E_1$ is also lower for dissociation at the step at Ni(211) than on Ni(111). The zero of energy in the plot corresponds to desorption of $NH_3$ from the surface. The $E_1$ for Ni(111) is thus 0.23 eV higher than $E_{ad}$ which means that the molecules would rather desorb than dissociate. At the step on Ni(211), $E_1$ just slightly exceeds $E_{ad}$ and dissociation and desorption are expected to be competitive, as observed in experiment [9] on Ru surface on which 70% of the $NH_3$ molecules desorb and only 30% of them dissociate. Since we find the $E_1/E_{ad}$ value to be much higher at the step edge than on Ni(111) terrace, we conclude that in the ammonia decomposition reaction the presence of the steps is important not only for recombinative desorption of nitrogen, but also for dissociation of $NH_3$. As already mentioned, we find the adsorption energy of $NH_3$ on Pd(111) to be even lower than on Ni(111). This result suggests that the reason for the low $NH_3$ decomposition rate on Pd [5] may be a low $E_{ad}/E_1$ ratio. To check this assumption we are calculating the barriers for $NH_3$ dissociation on Pd surfaces. The above results nevertheless show that although the $2N(ad) \Rightarrow N_2(gas)$ step in the $NH_3$ decomposition is rate-limiting, the $E_{ad}/E_1$ ratio is also important. If this ratio is substantially lower than 1, the system would not reach the $2N(ad) \Rightarrow N_2(gas)$ stage, because of desorption of molecules. Our results have already provided an important insight into the difference in the behavior on Ni and Pd surfaces toward $NH_3$ decomposition.

**Acknowledgment** We thank G. Ertl for helpful discussion. This work was suppoeted in part by DOE under grant No. DE-FGO3-03ER15464. Grant from NCSA, Urbana, IL was beneficial in providing computational resources.